\pdfoutput=1   
\documentclass{PoS}

\RequirePackage{lineno} 

\title{Energy measurement with the SDHCAL prototype}

\ShortTitle{Energy measurement with the SDHCAL prototype}

\author{\speaker{Alexey PETRUKHIN, for the CALICE Collaboration}

        IPNL/CNRS, France\\

        E-mail: \email{alexey.petrukhin@gmail.com}}

\abstract{The SDHCAL prototype that was completed in 2012 was exposed to beams of pions, electrons of different energies at the SPS of CERN for a total time period of 5 weeks. The data are being analyzed within the CALICE Collaboration. However preliminary results indicate that a highly granular hadronic calorimeter conceived for PFA application is also a powerful tool to separate pions from electrons. The SDHCAL provides also a very good resolution of hadronic showers energy measurement. A new calibration method that takes into account the degradation of the Glass Resistive Plate Chambers (GRPC) response for runs with rather  high particle beam rate is presented.}

\FullConference{Technology and Instrumentation in Particle Physics 2014\\

                 2-6 June, 2014\\

                 Amsterdam, the Netherlands}

\begin{document}


\section{Introduction}
The SDHCAL prototype  was conceived for two purposes. The first one is to confirm that  highly-granular gaseous hadronic calorimeters are capable of achieving  good resolution 
of the hadronic energy measurement while providing an excellent  tracking tool for the  Particle Flow Algorithms (PFA).
The second and most important aim is to demonstrate that such calorimeters are compatible with the requirements of  future ILC experiments in terms of efficiency, compactness and low power consumption.

In order to validate the SDHCAL technology, the prototype was exposed to muons, pions and electrons of the CERN H6 beam line of the SPS in September 2012, and of the H2 beam line in November 2012. We show here reanalysis of the same set of events collected during the September 2012 campaign and presented in the CALICE Collaboration analysis note CAN--037~\cite{CAN-37} and in the associated addendum-1. Also we show here the results  of the new analysis of the data taken during November 2012 runs. In both cases  to avoid efficiency loss in the GRPC in case of high particle rate only runs with less than 1000 particles per spill were studied.

\section{Prototype description}

\begin{figure}[htp]
\begin{center}
\includegraphics[width=0.75\textwidth]{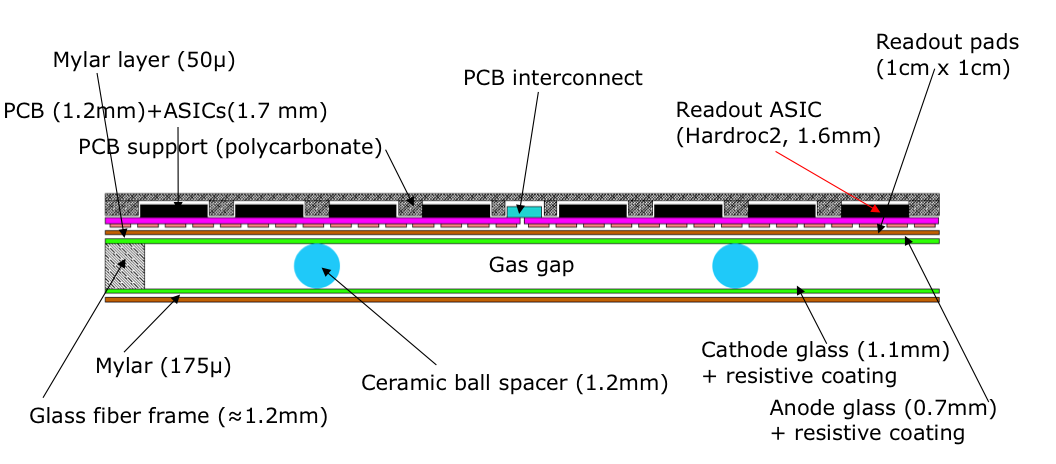}
\caption{A schematic cross-section of the active layer inside a SDHCAL cassette.}
\label{fig.scheme_GRPC}
\end{center}
\end{figure}

The SDHCAL comprises 48 active layers. Each of these layers is made of 1 m$^2$ Glass Resistive Plate Chamber (GRPC).  The GRPC signal  is  read out through 9216 pads of 1~cm$^2$ each. 
The pads are located on one face of an electronics board which hosts 144 HARDROC ASICs~\cite{hardroc} on its other side. 
The GRPC and the electronics board are put inside  a cassette made of two  stainless steel walls of 2.5~mm thickness each.  
The cassette keeps the pick-up pads of the electronics board in contact with the GRPC, and, it constitutes a part of the calorimeter absorber.  
 The total thickness of a cassette is 11~mm of which 6~mm are the active layer thickness occupied by the GRPC (3~mm), and the readout electronics (3~mm). 
 A cross-section of the active layer inside the cassette is shown in Figure~\ref{fig.scheme_GRPC}.

 The upper part of the cassette hosts also three Detector InterFace (DIF) cards which transfer the acquisition commands received through HDMI cables  to the ASICs of each slab, 
and collect the data received from these ASICs before forwarding the data through USB protocol to the acquisition stations.
 
 The 48 cassettes are then inserted into a self-supporting mechanical structure. The structure is built using 1.5~cm thickness stainless steel plates with a distance of 13 mm between 
two consecutive plates to allow an easy insertion of the cassettes. 
 The  acquisition mode used in the test beam (TB) was the triggerless mode. In this mode, data are collected after an acquisition command is sent to the ASIC. 
When  the memory of one ASIC is full, a RamFull command is sent to all, and the acquisition is stopped to allow the readout of the data recorded in the different ASICs. 
The acquisition restarts automatically upon the completion of the data transfer. During the data transfer no data is collected. This  dead time was reduced by increasing 
the number of USB buses for the data transfer. 
 
 The heating due to the power consumption  of more than 440 000 channels of the prototype  leads ineluctably to an increase of the prototype temperature which results in a change of the 
GRPC gain and an increase of the noise.  To avoid these problems the power-pulsing mode was used. This mode allows to keep the electronics in an idle mode during the time period separating 
two beam spills. In the case of the SPS beam cycle  this amounts to a reduction factor of five of the ASIC consumption (about nine seconds spill duration within a cycle of approximately 45 s).

\section{Data collection and quality control}

An important feature of the SDHCAL readout is the presence of three thresholds. The aim of using the thresholds information is not to measure the energy deposit in each pad but an attempt 
to distinguish between pads crossed by few, many or too many charged particles.  Information of three thresholds is coded in two bits. 
The thresholds values were fixed to 114~fC, 5~pC and 15~pC respectively, the average MIP induced charge being around 1.2~pC. The choice of these values was motivated by simulation studies.
 
No gain correction was  applied. The same electronics  gain  was used for all the channels (g=1). 

The gas mixture used to run the GRPC was made of  TetraFluoroEthane (TFE, 93{\%}),  $\mathrm{CO_2}$ (5{\%}) and $\mathrm{SF_6}$ (2{\%}). The high voltage applied on the GRPC was of 6.9 kV.

\begin{figure}[htp]
\begin{center}
\includegraphics[width=0.4\textwidth]{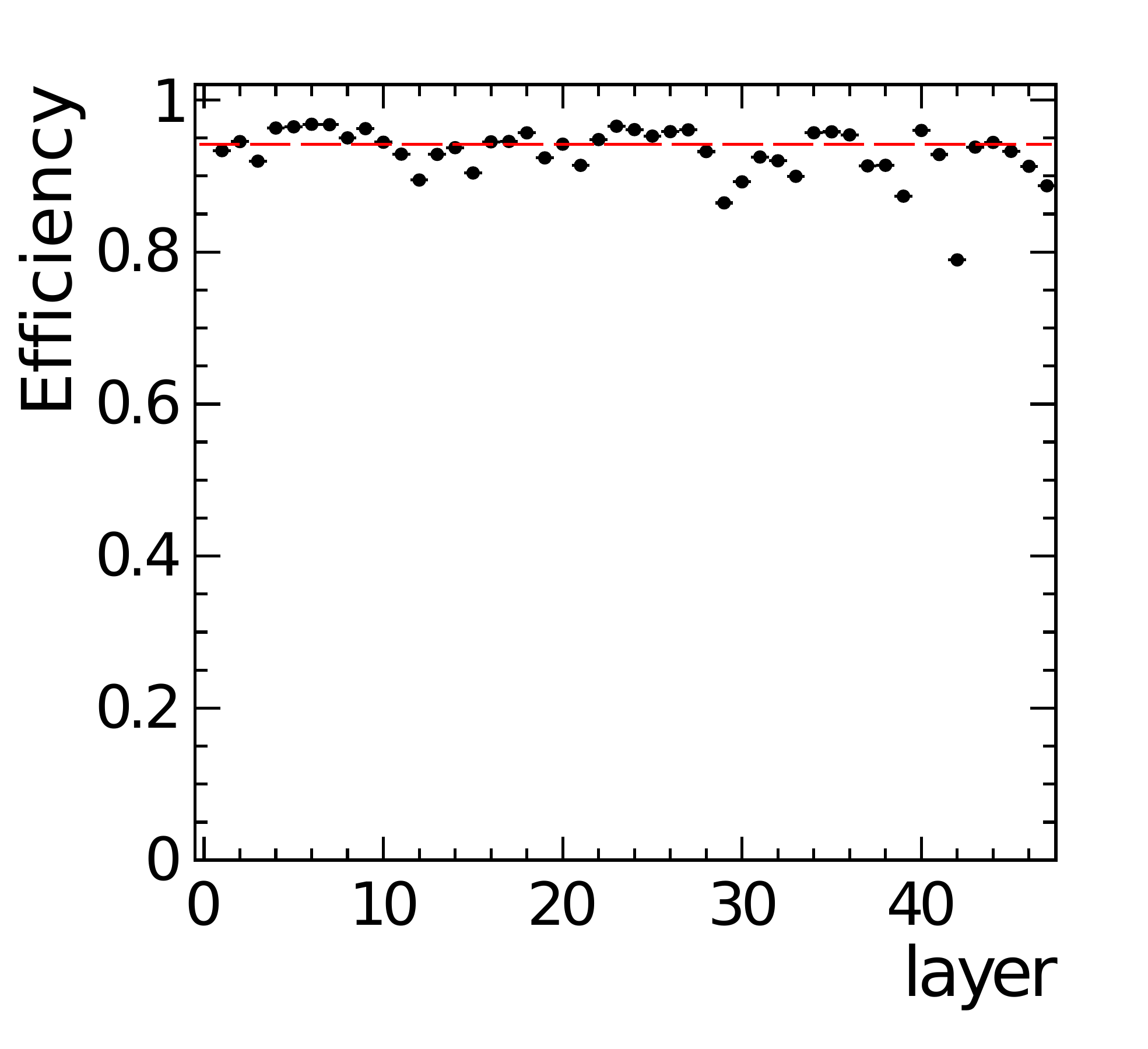}
\includegraphics[width=0.4\textwidth]{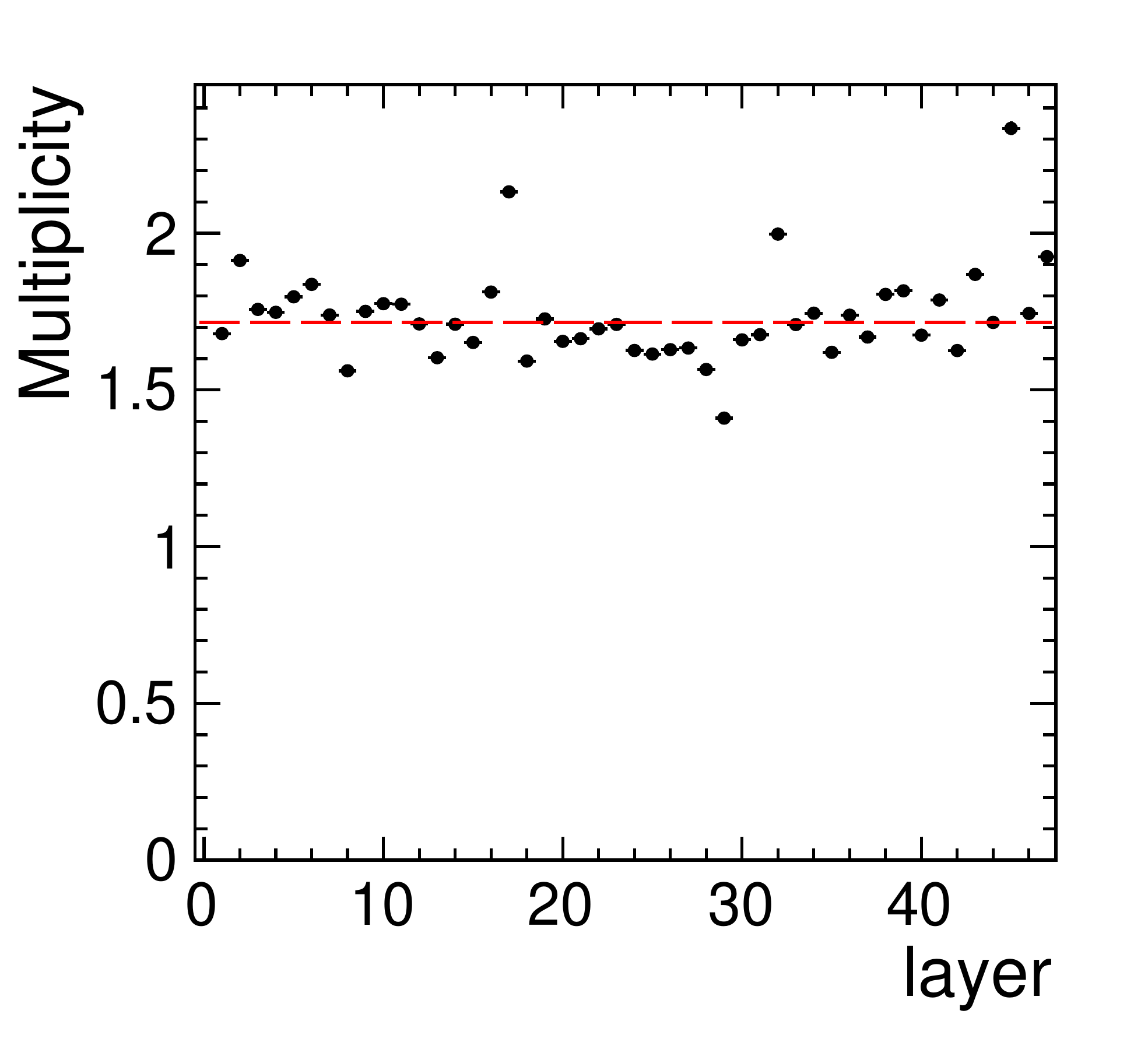}
\caption{Efficiency (left) and particle multiplicity (right) in September run. 
The red line is the average efficiency (left) and average multiplicity (right).
}
\label{fig.eff_august}
\end{center}
\end{figure}

To monitor the calorimeter performance, the efficiency and particle multiplicity of each of the 48 layers are estimated using the beam muons. To study the efficiency of one layer, tracks are used from the hits of the other layers. The expected impact point of the track in the layer under study is determined. The efficiency is then estimated as being the fraction of tracks for which at least one hit is found at a distance of less than 3~cm around the expected position. 
A track by track multiplicity is also estimated by counting the number of hits, if any, in the cluster built around the closest hit to the track's impact. A particle multiplicity for one sensitive region of the detector is then computed by averaging the track by track multiplicity for tracks going through 
the sensitive region under study. Figure~\ref{fig.eff_august} shows the efficiency and particle multiplicity of the layers during the September 2012 run.
Other methods to estimate the efficiency and particle multiplicity were performed, and confirm the results presented here.  

\section{Hadronic showers selection}

The original selection of pion sample is peresented in the CALICE Collaboration analysis note CAN--037~\cite{CAN-37}. We are showing here an extended version of that leading for the better rejection of electrons, muons and neutral particles in our data samples. 

\begin{figure}[!h]
\begin{center}
\includegraphics[width=0.4\textwidth]{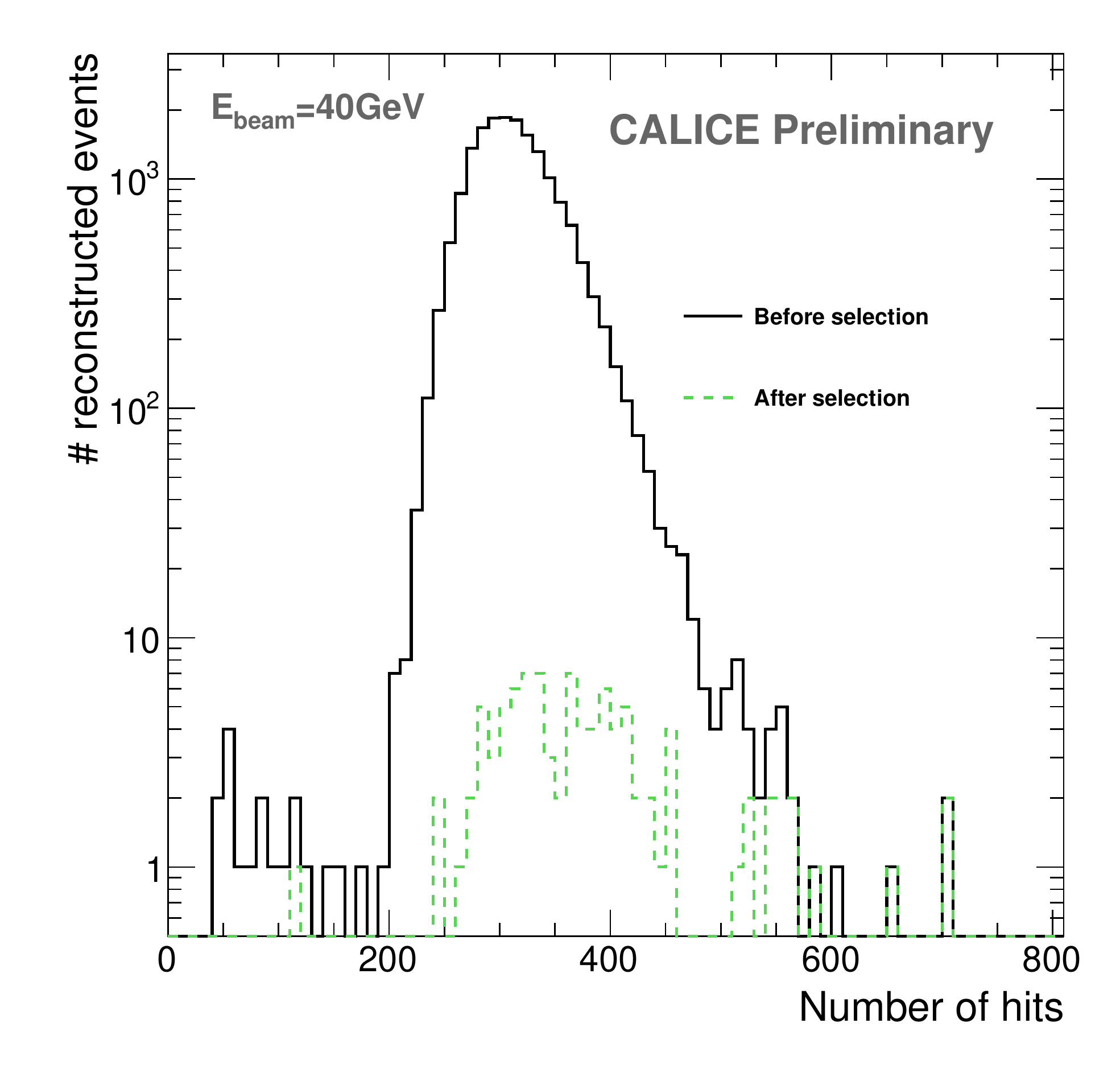}
\includegraphics[width=0.4\textwidth]{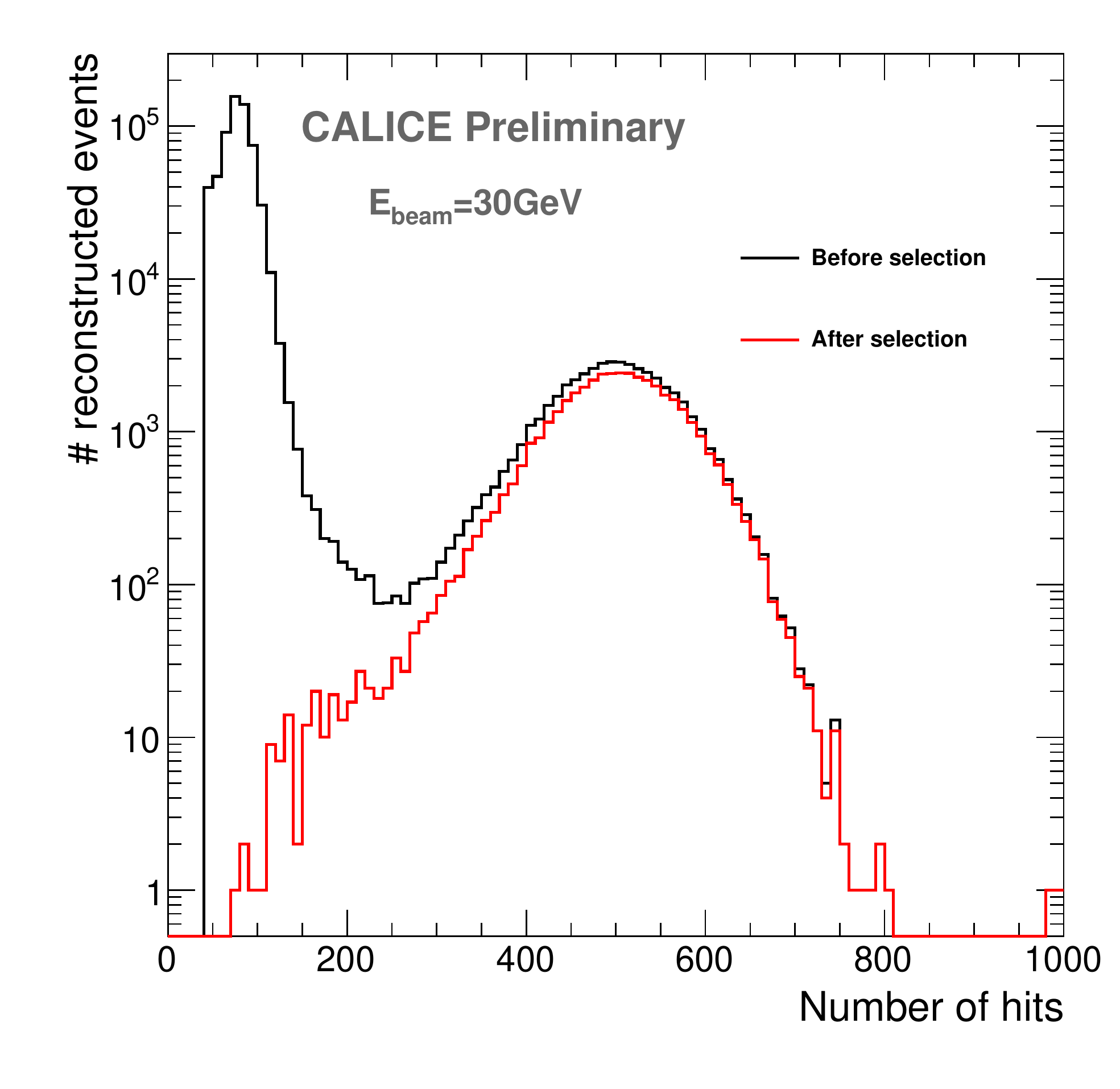}
\caption{Left: distribution of number of hits for 40 GeV electron run before (solid black line) and after (dashed green line) electron rejection. Right: number of hits for 30 GeV pion run before (black line) and after (red line) full selection.}
\label{fig:elec-rejection}
\end{center}
\end{figure}

Electrons are present in the pion beam despite the use of a lead filter to reduce their number. The absence of a Cherenkov counter or any other detector able to discriminate  electrons against pions makes it necessary to find other means to eliminate the electrons in our hadronic sample.  One way is to use the fact that electrons start their electromagnetic shower in the prototype in the first plates. This is due to the fact that the radiation length in steel is 1.76 cm.
For data which feature electromagnetic or hadronic showers, requiring that the shower starts in the fifth layer or after should in principle kill almost all of the electrons. To define the start of the shower we look for the first layer with more than 4 fired pads. To eliminate fake shower starts due to accidental noise or local high multiplicity effect, three consecutive layers were required to have more than 4 fired pads as well. 
Electromagnetic showers in the energy range between 5 and 80 GeV are longitudinally contained in less than 30 layers of our SDHCAL prototype. This electron rejection criterion was therefore applied only for events in which no more than 30 layers containing each more than 4 fired pads are found.  This limitation helps to minimize the loss of true pion hadronic showers at high energy where the number of fired layers exceeds 30. In this way high energy  pions starting  their shower in the first layers are not rejected.
Low energy pions have relatively smaller number of fired layers and are fully contained in the SDHCAL. For these pions the effect of this selection is only statistical. Pions that start showering in the first 4 layers are lost  but no  bias on the energy resolution is introduced.
Figure~\ref{fig:elec-rejection} (left) shows the distribution of number of hits before and after the selection for 40 GeV electron run. It shows clearly the rejection power of this selection.

Muons are also present in the pion beam. They are produced by pions stopped in the collimator or those decaying before reaching the prototype. To eliminate these muons as well as the cosmic contamination the average number of hits per fired layer was requested to be greater than $2.2$. This is higher than the average pad multiplicity which was found to be $1.73$~\cite{CAN-37}.  

To eliminate neutral contamination in our selection, events with strictly less than 4 hits in the first 5 layers are not considered. This criterion eliminates most of the cosmics as well.   
In addition to the previous criteria and in order to avoid the presence of more than one incoming particle in the final sample, events for which the first five layers have hits separated by more than 5 cm are eliminated. The result of such selection is shown in Figure~\ref{fig:elec-rejection} (right) for 30 GeV pion run where the total number of hits of the collected events is drawn. The fact that the shape of pion component is not affected ensures that our selection does not bias the hit distribution of  pions and hence the energy resolution should not be affected by this selection.

\section{Energy reconstruction}

An extension of the procedure presented in \cite{CAN-37} is used to determine the reconstructed energy of hadronic showers. In the extended procedure the energy is given by the following equation:

\begin{center}
\begin{equation}
  E_{reco}=\alpha (N_{tot}) N_{1} + \beta (N_{tot}) N_{2} + \gamma (N_{tot}) N_{3} + c N_{HT}
\end{equation}
\end{center}
\noindent
where  $N_{HT}$ is the number of hits belonging to the segments of the hadronic shower selected using the Hough Transform method as explained in \cite{CAN-47}. $N_{i}$ are the number of remaining hits associated to the $i^{th}$ threshold. $\alpha ,\beta, \gamma$, are quadratic functions of the total number of hits $N_{tot}$ and $c$ is a constant coefficient that reflects the fact that the HT segments are essentially produced by mips. The presence of high  thresholds in these segments is either a fluctuation or the result of large dE/dx at the stopping end and in both cases this has not the same signification as the thresholds associated to the hits present in the dense part of the shower. Therefore all the hits belonging to such segments are given the same weight.   The ten parameters are optimized using a part of September data of only few energy points.  The coefficients are obtained from a $\chi^2$ minimization using some of the energy bins:

\begin{center}
\begin{equation}
  \chi^2=\sum_{i=1}^{N}{\frac{(E_{beam}^i-E_{reco}^i)^2}{E_{beam}^i}}
\end{equation}
\end{center}

These coefficients are then used to estimate the energy of incoming particles.  The reconstructed energy distributions were fitted with a two-step Gaussian fit. First, a Gaussian was used to fit over the full range of the distribution. Second, a Gaussian was fitted only in the range of $\pm1.5 \sigma$ of the first fit. The $\sigma$ of second fit was used for the energy resolution estimation. The Crystal Ball function fit defined in \cite{CAN-37} was also performed. The difference of the two fits is used as the systematic error. 

\section{Spill time correction}

\begin{figure}[!h]
\begin{center}
\includegraphics[width=0.4\textwidth]{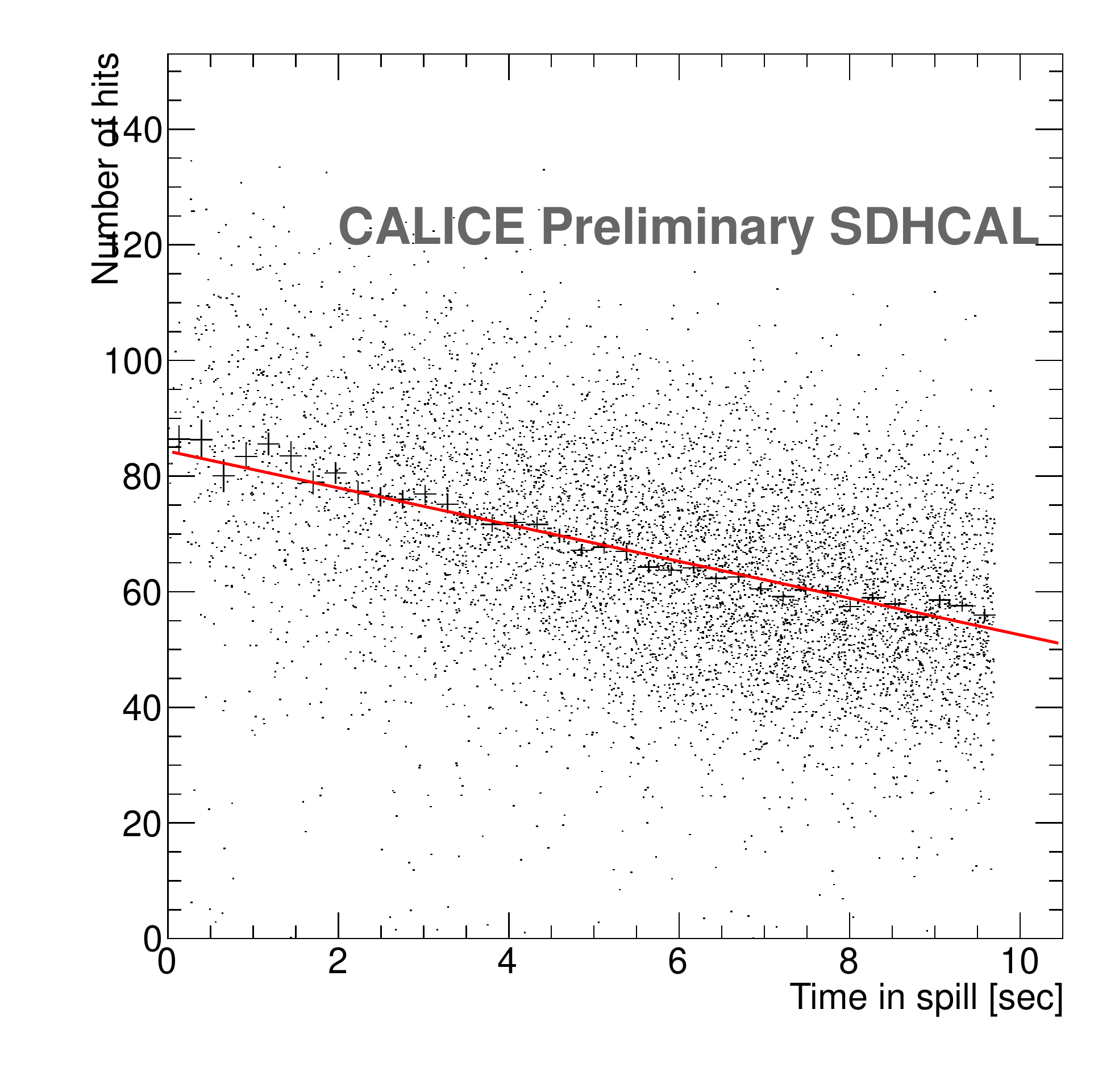}
\includegraphics[width=0.4\textwidth]{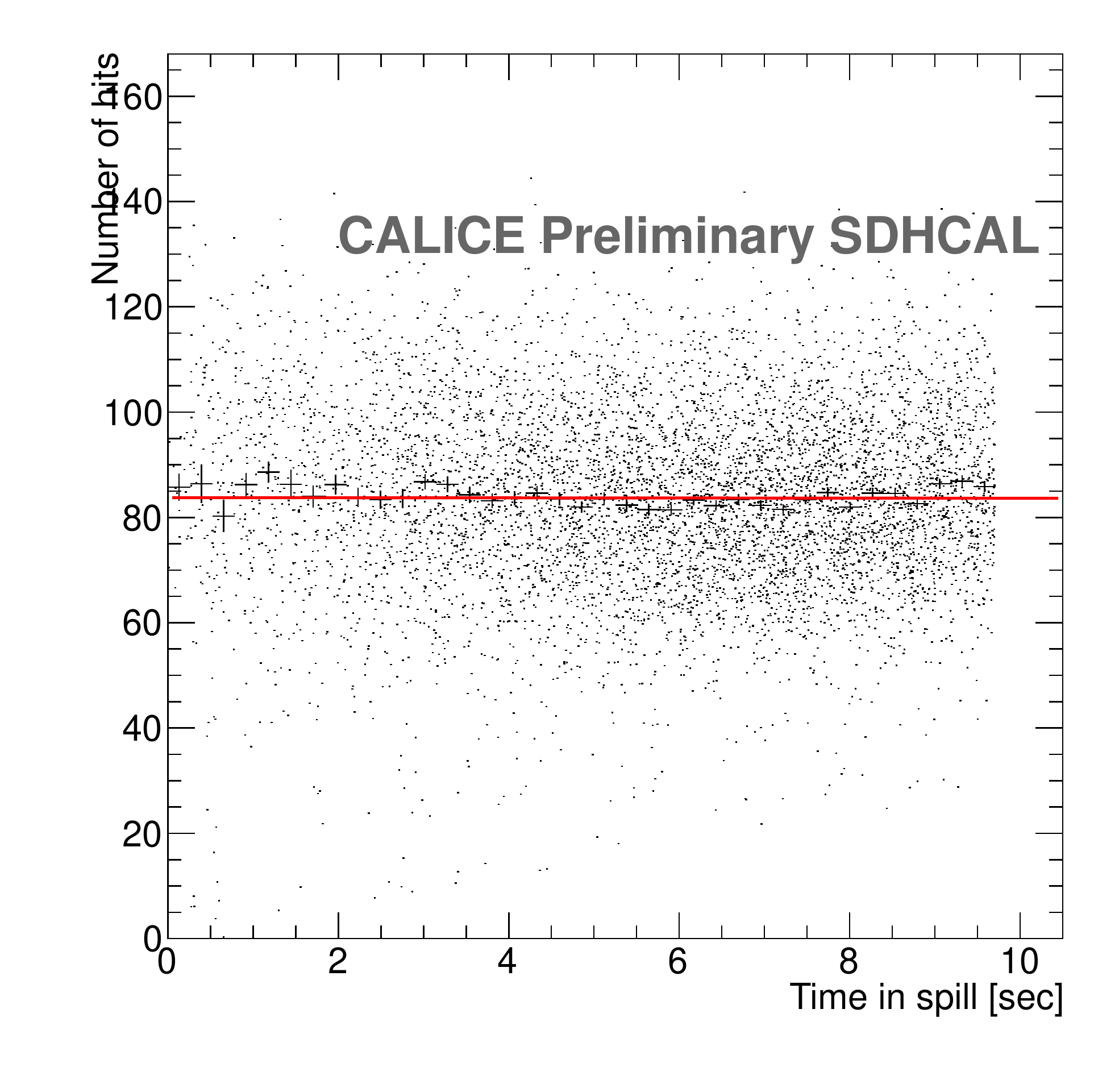}
\caption{Number of hits for the third threshold 80 GeV run as a function of spill time before (left) and after (right) linear fit calibration.}
\label{fig:linfit}
\end{center}
\end{figure}

Even though the beam parameters during the two data  taking periods were optimized to get spills containing less than 1000 particles it was observed that for some runs of both periods the number of hits associated to hadronic showers was decreasing during the spill time.  The decrease is more apparent for the number of hits associated to the second and third thresholds of the semi-digital readout.  The effect is more frequent  in runs of high energy pions. The consequence of such behavior is a degradation of  hadronic showers energy resolution.   In order to correct for the effect, two special calibration techniques were developed. The first one is a linear fit calibration. The average number of hits associated to each threshold of each hadronic shower  is plotted as a function of  their time occurrence within a spill. Then a linear fit to the hit distributions is performed and the slope of the fit is determined. The corrected number of hits $N_{corr}$ for each run is defined according to the following  formula: 

\begin{center}
\begin{equation}
  N_{corr}=\sum_{i=1}^{3}{Nhit_i - slope_i * TimeInSpill}
\end{equation}
\end{center}    

\noindent
where  $Nhit_i$ is the number of hits of a given threshold $i$ before correction and $TimeInSpill$ is the occurrence time within the spill.  The results before and after the linear fit calibration for 80 GeV run from September data can be seen in Figure~\ref{fig:linfit}. The alternative way of doing the correction is a time slots calibration. For each run and each threshold, the spill time was divided by 5 slots. Then a gaussian fit was performed for each of the number of hits distribution of each threshold for each time slot separately. The mean value from the fit for the first distribution (at the beginning of the spill) was taken as a reference. The correction factors for other 4 time slots are defined as $coeff_i=mean_1/mean_i$. The corrected number of hits $N_{corr}$ for each threshold is then defined as following: $N_{corr}=\sum_{i=1}^{5}{Nhit_i * coeff_i}$.

Both types of calibration are able to correct for the spill time effect. We observed that the energy resolution is slightly better for linear fit calibration however the linearity was found to be a little worse in this case. Finally the lack of statistics for some runs led us to choose the  linear fit calibration as the default one for both September and November data samples. 

 \section{Results}

\begin{figure}[!h]
\begin{center}
\includegraphics[width=0.49\textwidth]{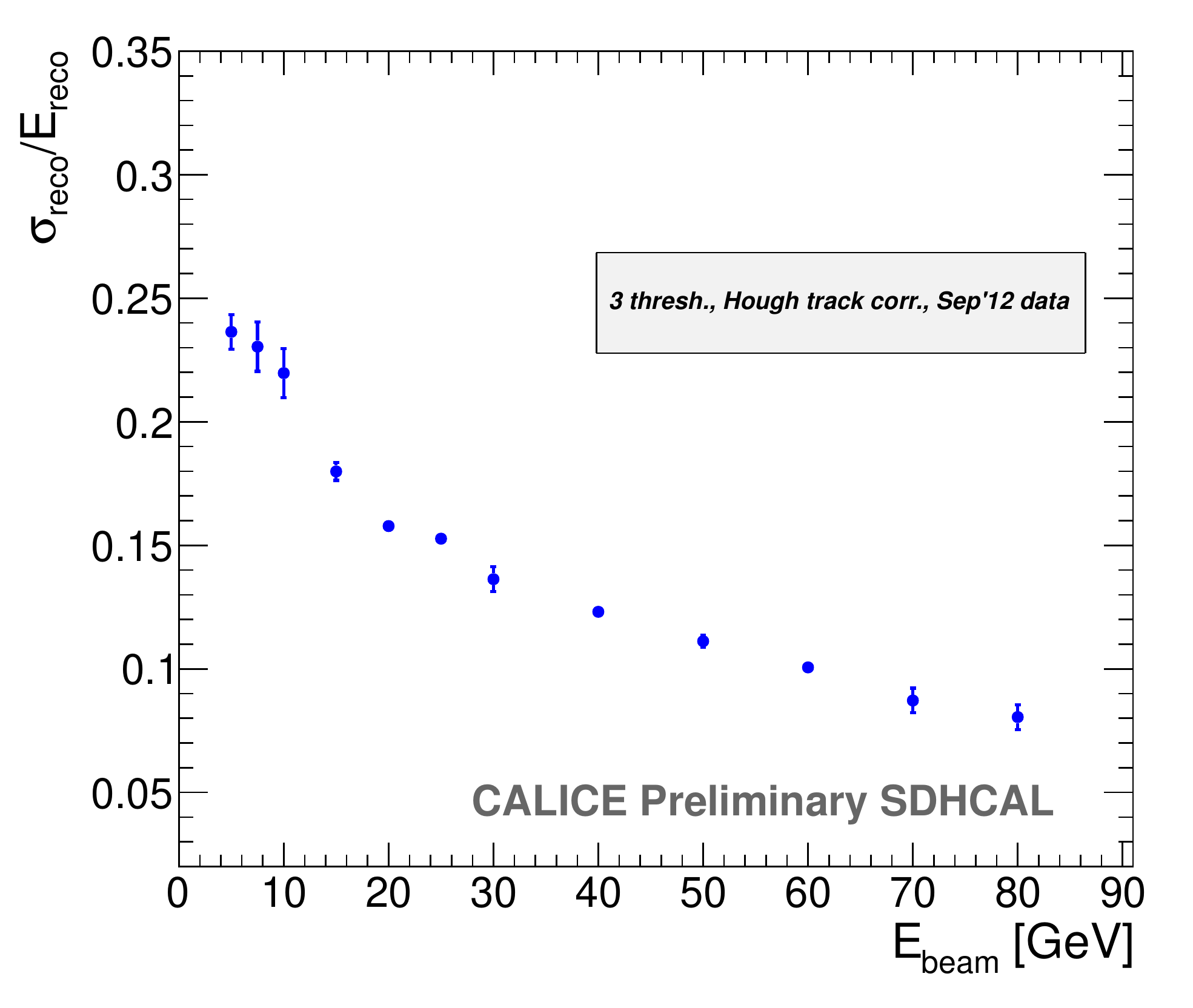}
\includegraphics[width=0.49\textwidth]{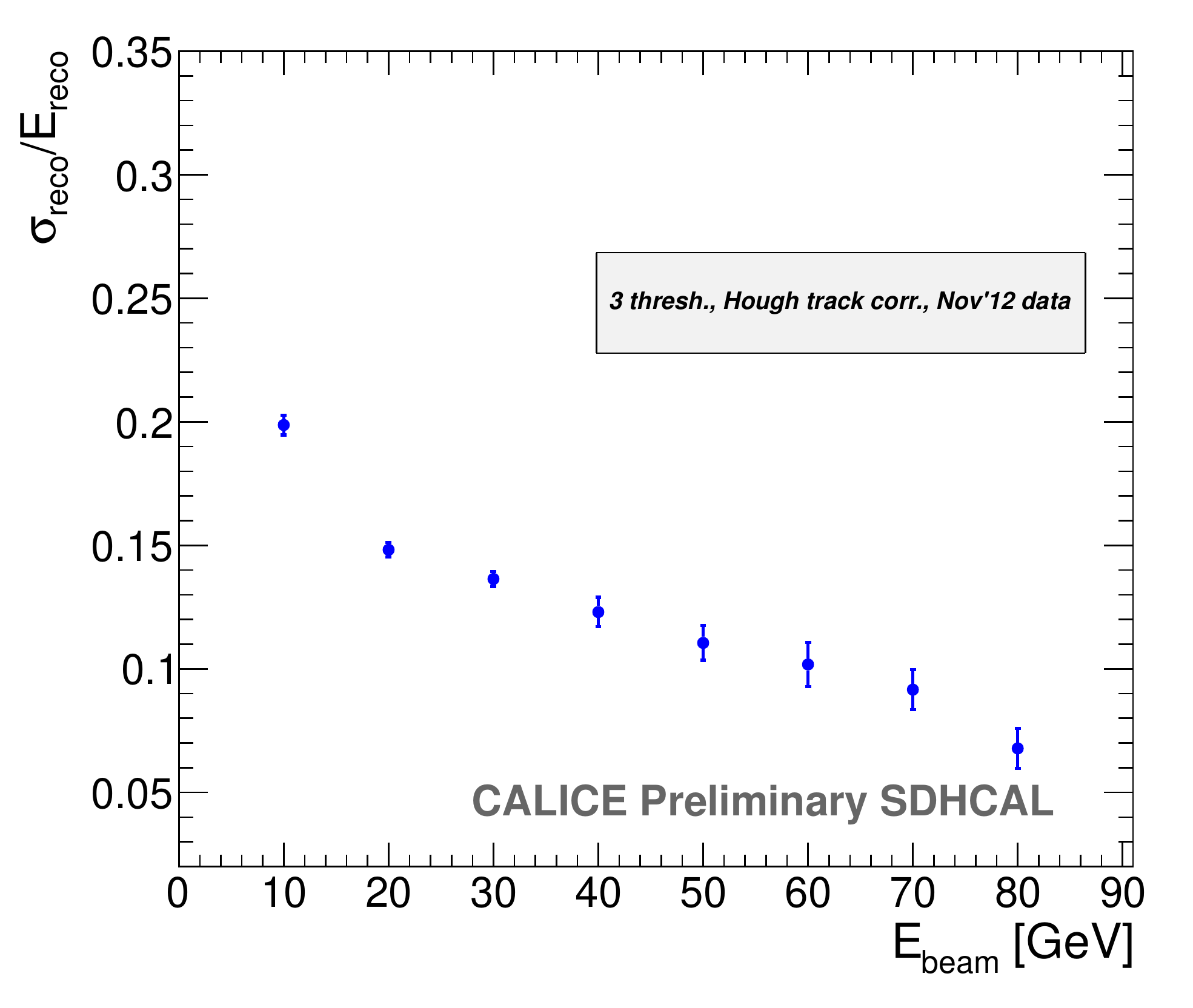}
\caption{$\frac{\sigma E_{reco}}{E_{reco}}$ of the reconstructed pion energy as a function of the beam energy at September (left) and November (right) runs.}
\label{fig:resolution}
\end{center}
\end{figure}

\begin{figure}[!h]
\begin{center}
\includegraphics[width=0.49\textwidth]{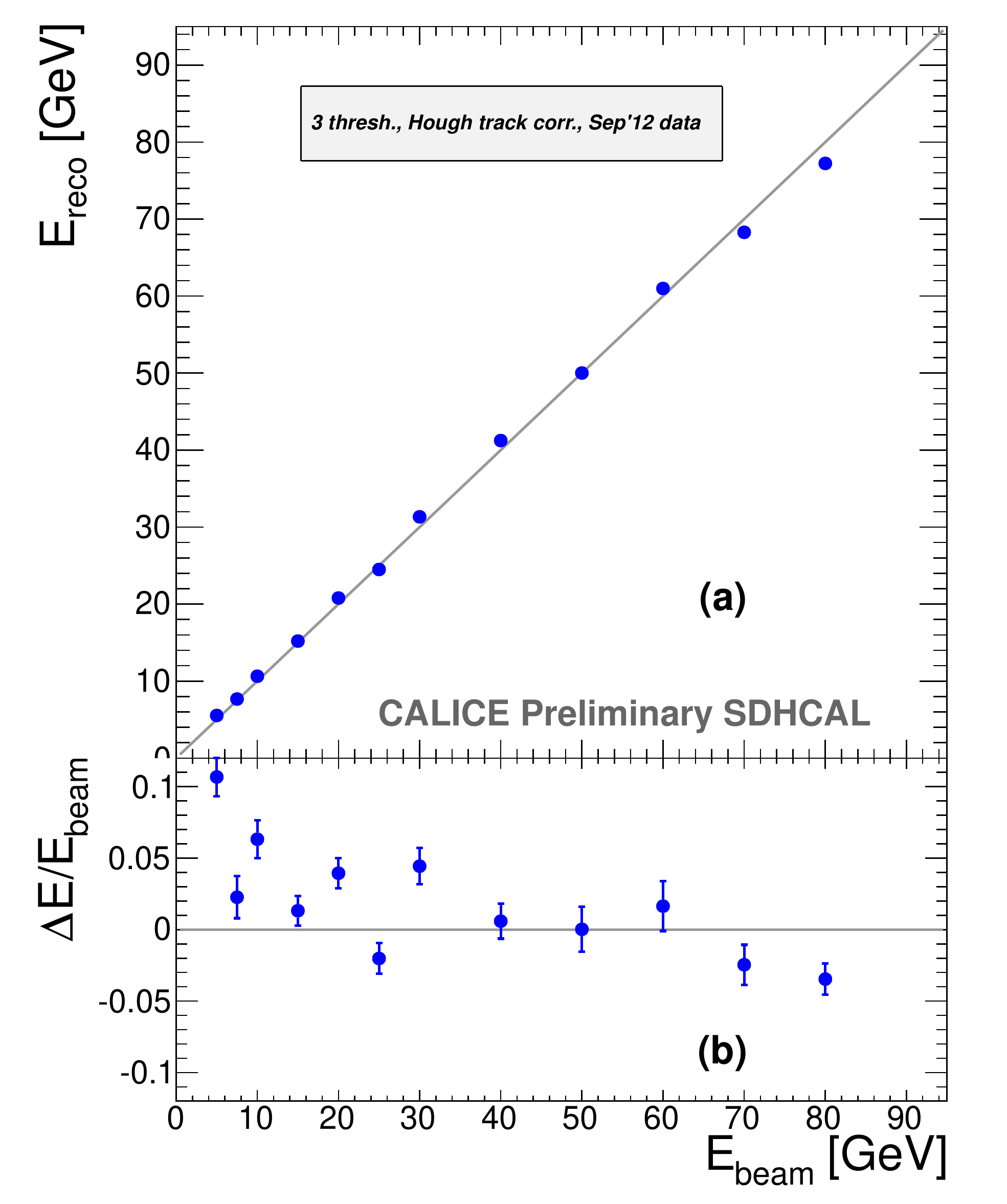}
\includegraphics[width=0.49\textwidth]{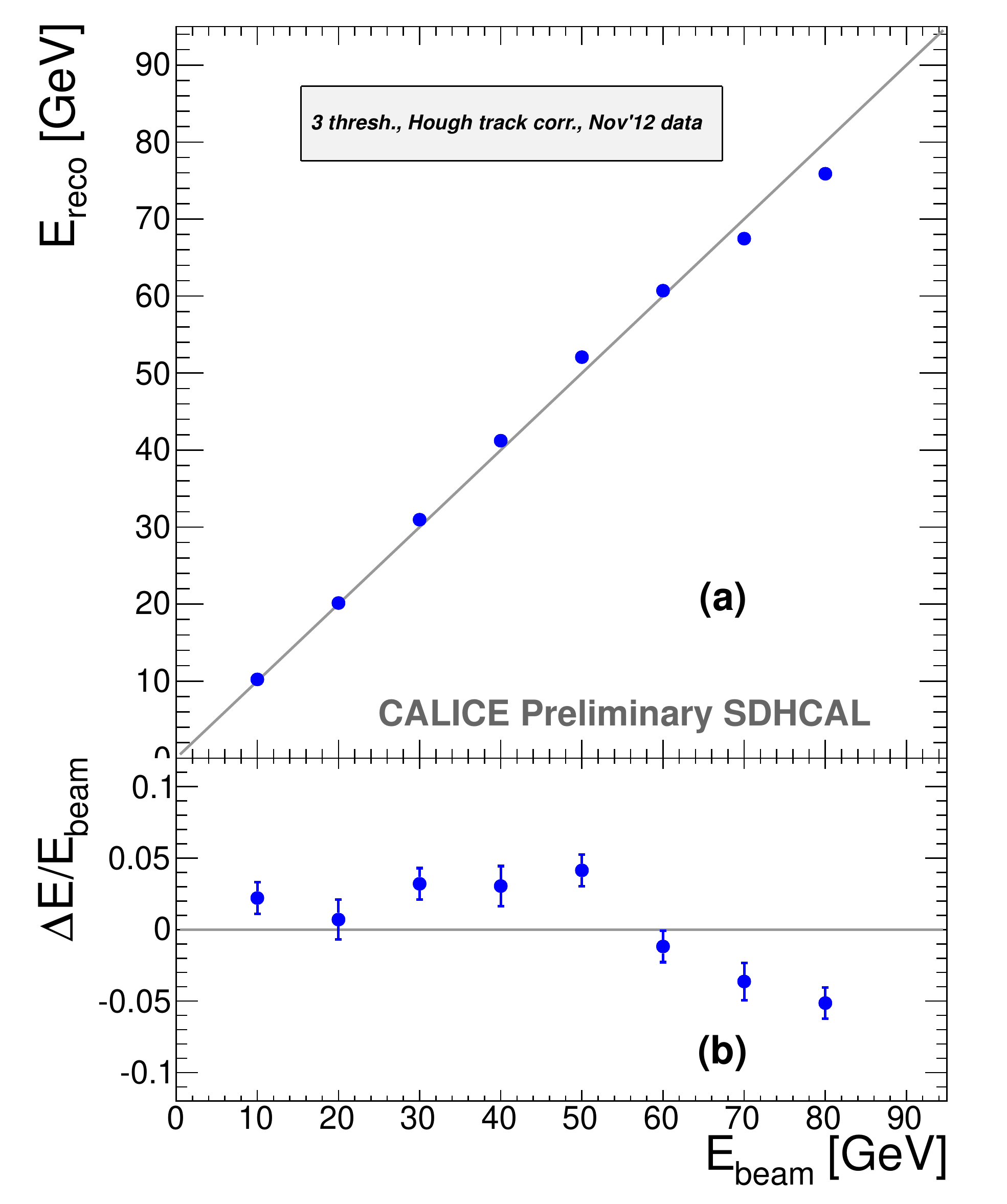}
\caption{(a): Mean reconstructed energy for pion showers at September (left) and November (right) runs; (b): relative deviation of the pion mean reconstructed energy with respect to the beam energy as a function of the beam energy at September (left) and November (right) runs.}
\label{fig:linearity}
\end{center}
\end{figure}

The energy resolution and linearity of the two sets of data are presented in Figures \ref{fig:resolution} and \ref{fig:linearity}. The improvement  of the September  data 
 with respect to the results presented in addendum-1 to CAN--037~\cite{CAN-37} is obvious at high energy. This improvement in energy resolution is as high as 20\% in some cases. The value of resolution reaches 6.8\% at 80 GeV for November data.

The parameters used for energy reconstruction were optimized with September data set only. Application of those parameters to November data set (where beam conditions are different) shows a good agreement between two data periods. It clearly demonstrates that the behavior of  SDHCAL prototype was stable between the two periods.

\end{document}